\documentclass[review]{elsarticle}

\usepackage{natbib}

\usepackage{longtable}

\journal{Journal of \LaTeX\ Templates}

\begin{document}

\begin{frontmatter}

\title{Orbital Period Changes in IO~Cep, IM~Cep and TX~Ari: Path to Masses of Distant Components}

\author{V. Bak{\i}\c{s}\corref{mycorrespondingauthor}}
%\fntext[myfootnote]{email:volkanbakis@akdeniz.edu.tr}
\ead{volkanbakis@akdeniz.edu.tr}

\author{Z. Eker, H. Bak{\i}\c{s}, S. Kayac{\i}, G. Y\"{u}cel, E. Tun\c{c}, \"{O}. Ta\c{s}p{\i}nar, Y. Yal\c{c}{\i}n, A. Melnik, \c{C}. Esenda\u{g}l{\i}}

\address{Akdeniz University, Department of Space Sciences and Technologies, Dumlupinar Blv., Kamp\"{u}s, 07058, Antalya, TURKIYE}

\begin{abstract}
Three Algol-type binary systems (IO~Cep, IM~Cep and TX~Ari) showing cyclic orbital period changes are studied. The combination of time of minimum data from the ground-based observations together with high precision photometric data from the TESS satellite enabled us to estimate the basic light curve elements of binary systems and mass functions for distant components around the systems. The relation of mass ratio to the system geometry in semi-detached binary stars allowed us to determine the mass ratio of the binary components without using spectra. By using the color and distance information from the GAIA EDR3 and light contributions of the components from the light curve analysis, the astrophysical parameters of the binary components as well as the minimum masses of the distant components are obtained with an uncertainty of $\sim$10-20 per cent indicating that the method can be a good guide for those studying with faint systems where spectra with sufficient resolution and S/N ratio are difficult to acquire.
\end{abstract}


\end{frontmatter}

%\linenumbers

\section{Introduction}
\label{sec:intro}

A perturbation on the mid-eclipse timing of eclipsing binaries due to an external body is known as Light Travel Time (LTT) effect. LTT effect is a powerful tool to investigate distant components in eclipsing systems  where an eclipsing pair moves about the center of mass of a wider triple system, (a lesser chance a quadruple system) which causes a periodic variations in the mid-eclipse times of the eclipsing binary. Eclipses behave as an accurate clock in detecting distance variations of the eclipsing pair from the center of mass of the triple (or quadruple) system. Who possibly inspired by Ole Roomer’s calculation (1676) on the speed of light using LTT effect on Io and Jupiter, \cite{Chandler} is first to discuss LTT effect as a possible mechanism to explain period changes of Algol. Formulation of the problem is given by \cite{Irwin1952} and \cite{Irwin1959}.

The LTT effect on $O-C$ diagrams was revised by \cite{Friboes1973}, \cite{Mayer1990} and \cite{Wolf2014}. Recently, the LTT effect was used for triple star investigations among the KEPLER binaries \citep{Rappaport2013}. There are many applications of LTT effect formulations, for example on HW Vir  systems (\cite{Lee2009}; \cite{Almeida2013}), on the eclipsing binaries of DA + dM (\cite{Parsons2010}: \cite{Beuermann2011}; \cite{Qian2012}), on Cataclysmic systems \citep{Dai2010} and RS CVn systems \citep{Liao2010}; \cite{Erdem2007} on Algol-type binaries. Study of \cite{Budding2005} and \cite{Bakis2006}, in which Hipparcos astrometry and LTT effect are analyzed together, is a good example of how it can be effectively used to access the mass of the third body. LTT effect could be useful even on detecting circum-binary planets \citep{Zorotovic2013} and black holes \citep{Qian2008,Wang2021}. We may count \cite{Almeida2019} and \cite{Ozavci2020} for the recent applications of LTT effect.

In this study, we are performing the period analysis of three semi-detached Algol-type systems for the first time (\S3). Period analysis indicates that there should be at least one additional component in two systems and one system needs further evidence. We analyzed available photometric data from the TESS satellite \citep{tess} to estimate the minimum mass of the distant objects (\S4). Using the GAIA distance for each system, we arrived at the absolute parameters of the binaries and then the minimum mass of the additional component. At the end of the study (\S5), we discuss the minimum mass of the additional components.

\section{Observations and Data Reduction}
\label{sec:obs}

The observations of IO~Cep, IM~Cep and TX~Ari have been performed within the framework of a program where we systematically obtain the times of minima of the eclipsing binary stars using the remotely controlled AUT25 telescope of the Akdeniz University Space Sciences and Technologies Department. AUT25 telescope optical tube assembly (OTA) is a Ritchey-Chretien type with a primary mirror of 10-inch and the mounting system is EQ-8 model german-equatorial type. The telescope is equipped with a QSI-632ws CCD and Johnson UBVRI filters. The autoguiding is performed with a Orion StarShoot camera attached to a 80mm refractor which is also attached to the OTA. The autoguiding unit allows the observer to take CCD frames up to exposure time of 1200s which enables it to observe stars down to 18 magnitude in Johnson V-band.

The program stars have been observed between December 2017-December 2019 in different filters. The selection of the filter has been done on the basis of the brightness of the object during the eclipse in order to keep the S/N ratio as high as possible. The mid-eclipse time was measured using the Kwee-van Woerden method \citep{Kwee1956} and parabolic fitting to the centeral part of the minimum. The final time of minimum and its uncertainty have been determined by taking the weighted mean of the measurements. The measured times of minima, their uncertainties, types of minimum and the filters used for IO~Cep, IM~Cep and TX~Ari have been given in Table~\ref{tab:minima}.

\begin{center}
	\begin{longtable}{|l|c|c|c|}
		\caption{Observed times of minima for program stars. Errors in brackets refer to standard errors.} \label{tab:minima} \\
		
		\hline \multicolumn{1}{|c|}{\textbf{Star}} & \multicolumn{1}{c|}{\textbf{Time of Minimum}} & \multicolumn{1}{c|}{\textbf{Filter}} & \multicolumn{1}{c|}{\textbf{Pri/Sec}}\\
		\multicolumn{1}{|c|}{\textbf{ }} & \multicolumn{1}{c|}{\textbf{HJD-2400000}} & \multicolumn{1}{c|}{\textbf{ }} & \multicolumn{1}{c|}{\textbf{ }}\\ 
		\hline 
		\endfirsthead

		\multicolumn{4}{c}%
		{{\bfseries \tablename\ \thetable{} -- continued from previous page}} \\
		\hline \multicolumn{1}{|c|}{\textbf{Star}} & \multicolumn{1}{c|}{\textbf{Time of Minimum}} & \multicolumn{1}{c|}{\textbf{Filter}} & \multicolumn{1}{c|}{\textbf{Pri/Sec}}\\
		\multicolumn{1}{|c|}{\textbf{ }} & \multicolumn{1}{c|}{\textbf{HJD-2400000}} & \multicolumn{1}{c|}{\textbf{ }} & \multicolumn{1}{c|}{\textbf{ }}\\ \hline 
		\endhead

		\hline \multicolumn{4}{|r|}{{Continued on next page}} \\ \hline
		\endfoot

		\hline \hline
		\endlastfoot
		TX~Ari    & 58805.3145(8) & R & p \\
				  & 58821.4588(2) & R & p \\
		IM~Cep    & 58806.3069(3) & I & p \\
				  & 58824.2760(2) & I & s \\
		IO~Cep    & 58793.2328(3) & V & p \\
	\end{longtable}
\end{center}

\section{Orbital Period Analysis}
\label{sec:oc}

In addition to our observations in Table\ref{tab:minima}, available times of minima were collected from the literature and new times of minima were extracted from the TESS satellite photometric data, which are presented in electronic form at CDS\footnote{The time of minima are available via anonymous ftp to cdsarc.u-strasbg.fr(130.79.128.5)}. Two main database namely $O-C$ Gateway\footnote{http://var2.astro.cz/ocgate/} and the "An Atlas of $O-C$ Diagrams of Eclipsing Binary Stars" \cite{Kreiner2004} were used for the old and recent available times of minima.

The methodology of the data handling and analysis is the following: First, the data is weighted according to their observing techniques, namely visual observations (vis) are weighted with 1, photographic observations (pg) are weighted with 5 and CCD \& Photoelectric (ccd/pe) observations are weighted with 10. This kind of weighting is succesfully used in recent studies \citep[e.g.][]{Erdem2007} where the uncertainty of times of minima are not available. The $O-C$ residuals were calculated using the linear ephemeris ($T_\mathrm{0}$, $P$) which are taken mainly from \cite{Kreiner2004}. Since the linear ephmeris for IM~Cep does not exist in \cite{Kreiner2004}, it is adopted from the $O-C$ Gateway.

\begin{figure}[!t]
	\includegraphics[width=\columnwidth]{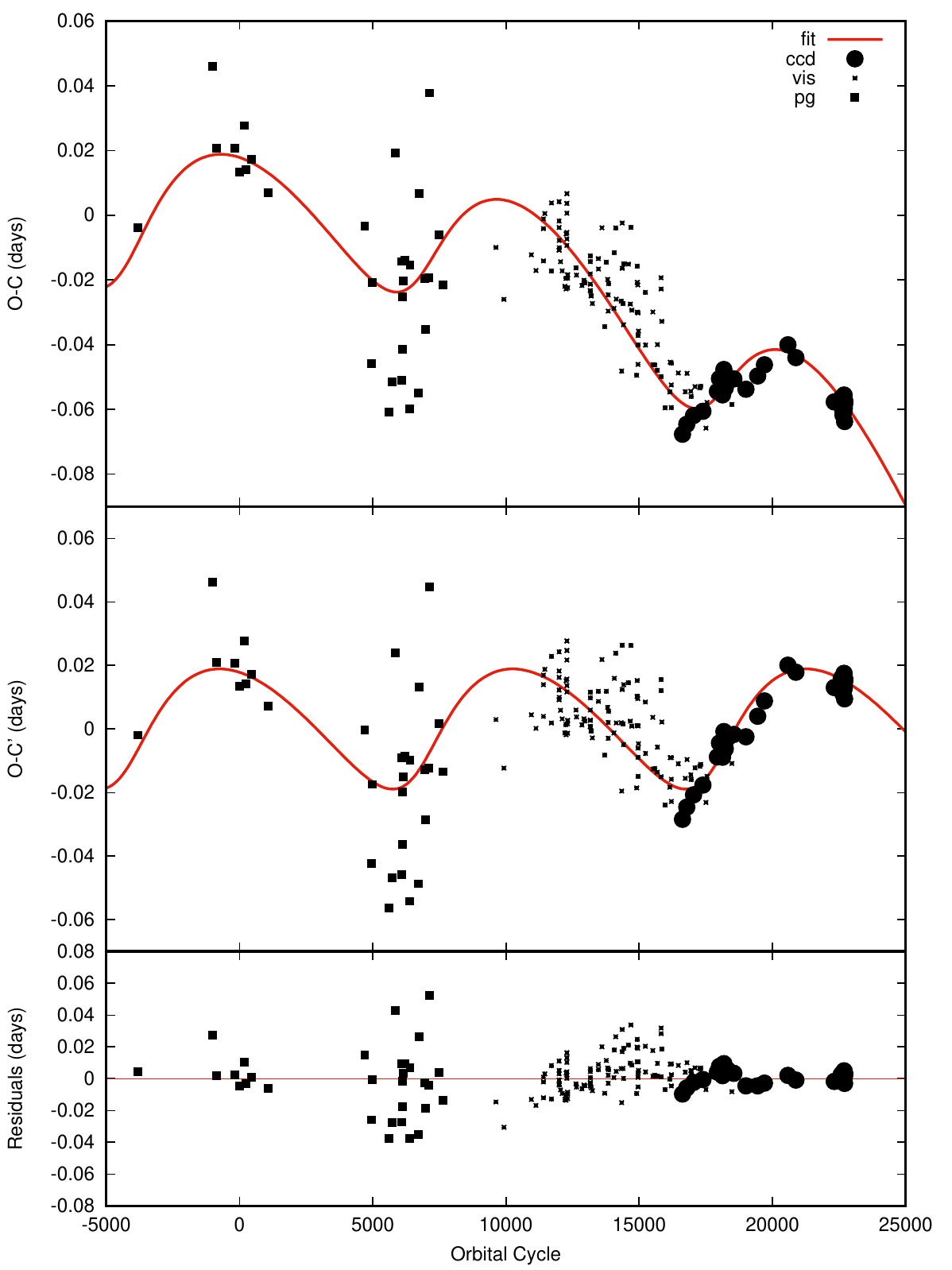}
	\caption{LTT fit on the $O-C$ curve of IO~Cep.}
	\label{fig:iocep}
\end{figure}

\begin{figure}[!t]
	\includegraphics[width=\columnwidth]{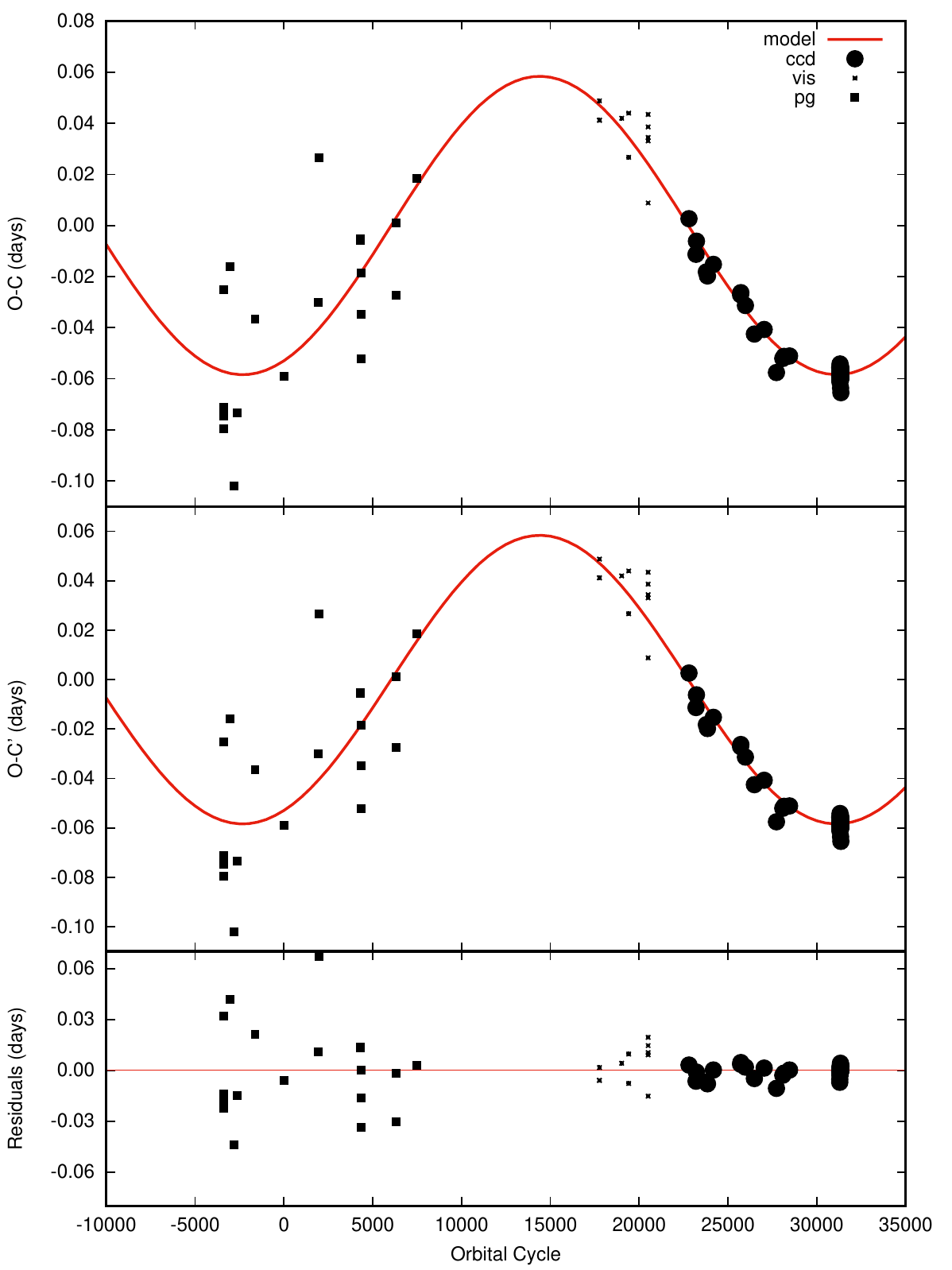}
	\caption{LTT fit on the $O-C$ curve of IM~Cep.}
	\label{fig:imcep}
\end{figure}

\begin{figure}[!t]
	\includegraphics[width=\columnwidth]{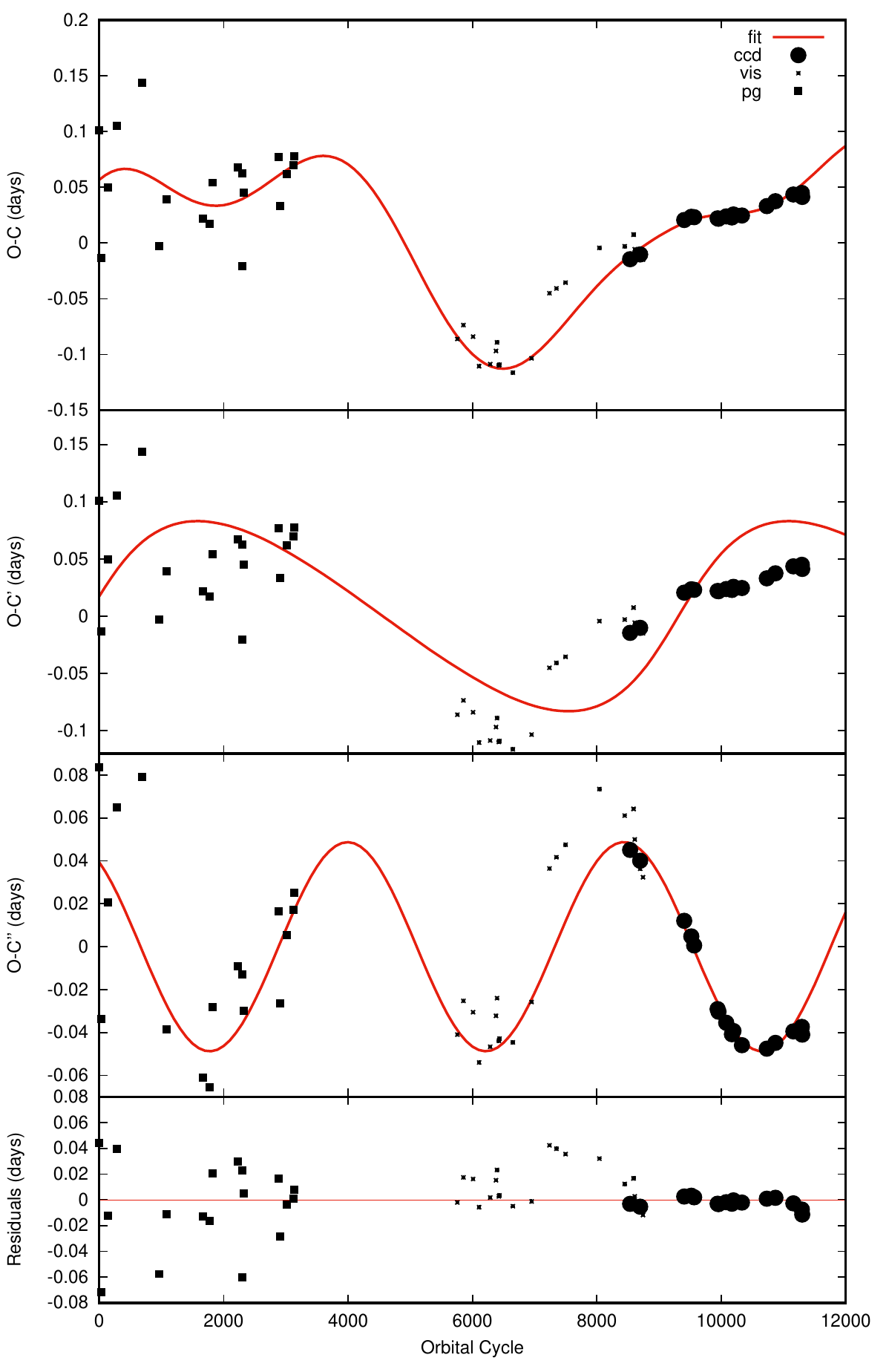}
	\caption{LTT fit on the $O-C$ curve of TX~Ari.}
	\label{fig:txari}
\end{figure}

As can be seen from the Figs.\ref{fig:iocep}-\ref{fig:txari}, all systems show cyclic variation in their $O-C$ diagram. Therefore, the $O-C$ data have been examined to find out whether LTT or magnetic activity play role in this cyclic period change. The time delay ($\Delta t$) due to the motion of the eclipsing binary around the common center of mass with the additional component(s) is first derived by \cite{Irwin1952} as in the Eq.~\ref{eq:irwin}.

\begin{equation}
	\label{eq:irwin}
	\Delta t = \frac{a_{12}sin i_{12}}{c}
	\left\{
	\frac{1-e^2_{12}}{1+cos\nu_{12}}sin(\nu_{12}+\omega_{12})+e_{12}cos\omega_{12}
	\right\}
\end{equation}

where $a_{12}$ is semi-major axis length, $i_{12}$ is inclination, $e_{12}$ is eccentricity, $w_{12}$ is longitude of periastron of the orbit, $\nu_{12}$ is true anomaly and $c$ is speed of light.

Together with the linear and parabolic terms, the $O-C$ data have been analyzed with Eq.~\ref{eq:analysis} which calculates the expected value of the time of minimum. The difference of observed times of minima from Eq.~\ref{eq:analysis} constructs the $O-C$ residuals.

\begin{equation}
	\label{eq:analysis}
	O - C = \Delta T_0 + \Delta P\times\,E + Q\times\,E^2 + \Delta\,t
\end{equation}

where $\Delta T_0$, $\Delta P$ and $\Delta t$ are the correction terms for reference time, orbital period and light travel time, respectively, $E$ is epoch number and $Q$ is the quadratic term which is used for estimating the mass loss and/or transfer in the system. In the case of conservative mass transfer, the $Q$ parameter is related to the amount of mass transfer rate between components as in Eq.\ref{eq:masstransfer}.

\begin{equation}
	\label{eq:masstransfer}
	\frac{\Delta P}{P}=\frac{2Q}{P}=3\Delta m_1\frac{m_1-m_2}{(m_1+m_2)^2}
\end{equation}

where $m_{1,2}$ are masses of the components and $\Delta m_1$ is the amount of the transferred mass from secondary to primary component.

As can be seen from the Eq.\ref{eq:masstransfer}, besides the Q-parameter, the total mass of the system must be known in order to determine the amount of mass transfer.

The mass function for the physically bound third body is given as in Eq.\ref{eq:massfunction}, which tells us that the total mass of the binary system must be known priori in order to calculate the minimum mass of the physically bound distant component. The methodology used to determine the total mass of the binary star is explained in detail in \S4.

\begin{equation}
	\label{eq:massfunction}
	f(m_3) = \frac{m^3_3 sin^3i}{(m_{12}+m_3)^2} = \frac{(a_{12} sin i)^3}{P_{12}^2}
\end{equation}

The third body orbital parameters obtained as a result of the application of the Eq.\ref{eq:analysis} are given in Table~\ref{tab:pars} and the model fits are shown in Figs.~\ref{fig:iocep}-\ref{fig:txari}. From Table~\ref{tab:pars}, it is seen that IM~Cep has a circular third-body orbit and TX~Ari has a circular fourth-body orbit implying a possiblity of magnetic activity cycle. The third light contribution in IM~Cep (see $\S$\ref{sec:lc_analysis}) confirms a distant component while TX~Ari requires multiband observations for further evidence.

\begin{table}[!t]\centering\small
	\caption{Orbital parameters for wide orbits. Errors quoted in brackets refer to standard errors in the last digit.} \label{tab:pars}
	\begin{tabular}{cccc}
		\hline
		Parameter              & IO~Cep          & IM~Cep          & TX~Ari   \\
		\hline
		$P_{12}$ (days)        & 1.2358111(6)    & 0.9215840       & 2.691325(1)    \\
		$T_{12}$ (HJD-2400000) & 30,729.2661(36) & 29,931.2983(78) & 28,409.445(37) \\
		$Q$ (x10$^{-10}$)      & --1.43(25)      & --0.11(3)       & 0.0            \\
		$T_3$ (HJD-2400000)    & 40,932(267)     & 35,500(791)     & 53,468(1584)   \\
		$P_3$ (days)           & 13,476(284)     & 30,812(1,078)   & 25,573(2,190) \\
		$P_3$ (yrs)     & 36.9(0.8) & 84.4(3.3)  & 70.0(6.0) \\
		$asini_{3}$ (AU)       & 3.4(3)          & 10.1(1.1)       & 14.4(2.9)     \\
		$e_3$                  & 0.32(13)        & 0.0             & 0.4           \\
		$w_3$ (rad)            & 0.0(6)          & 0.0             & 0.0           \\
		$P_{4}$ (days)         & -               & -               & 11,943(941)   \\
		$P_{4}$ (yrs)         & -               & -               & 32.7(2.6)   \\
		$asini_{4}$ (AU)       & -               & -               & 8.4(5.1)      \\
		$T_{4}$ (HJD-2400000)  & -               & -               & 48,393(870)   \\
		$e_4$                  & -               & -               & 0.0(0.3)      \\
		$w_4$ (rad)            & -               & -               & 0.1(0.2)      \\
		$f(m_3)$               & 0.029(8)        & 0.145(17)       & 0.609(10)     \\
		$f(m_4)$               & -               & -               & 0.563(125)    \\
		$\chi^2$               & 0.029(9)        & 0.0714          & 0.0670        \\
		\hline
	\end{tabular}
\end{table}

\section{Astrophysical Parameters}
\label{sec:lc_analysis}

The best way to find the absolute parameters of an eclipsing binary star is to analyze the photometric data and spectral data together. However, in most cases, especially in Algol-type semi-detached systems, it is not possible to see the spectral lines of both components in the combined spectrum due to the relatively less light contribution of the secondary component. In such cases, it may be a good option to use the mass-ratio ($q$) search method for the system's mass ratio. Because, the only parameter that determines Roche geometry is the mass-ratio of the components. In semi-detached systems, since the secondary component fills the Roche lobe, the radius of the secondary component is determined by the mass-ratio of components $q$ ($M_2$ / $M_1$). That is, $q$ is a parameter that alters the light curve. Thus, a $q-\chi^2$ graph can be obtained by looking at the goodness of model fits with different $q$ values.

\begin{figure*}
	\begin{center}
		\begin{tabular}{ccc}
			\resizebox{140mm}{!}{\includegraphics{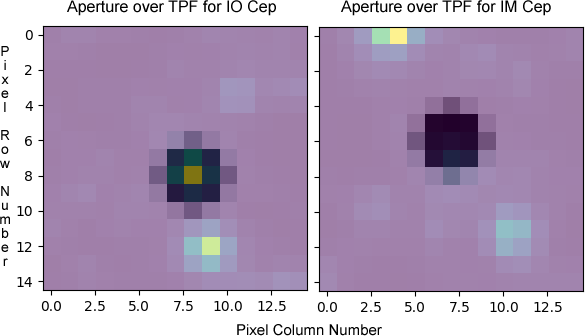}} & 
		\end{tabular}
		\caption{TESS aperture used for flux reading. The black area is the best aperture frame determined by the {\sc eleanor} \citep{Eleanor2019}.} \label{fig:tess_aperture}
	\end{center}
\end{figure*}

Except for the TX~Ari, systems studied in this article have been observed during the TESS (Transiting Exoplanet Survey Satellite) \cite{tess} mission in different sectors. We transformed the observations in these sectors into light curve data using the code {\sc eleanor} \citep{eleanor}. Star fields on the target pixel file (TPF) showing the best aperture values determined by {\sc eleanor} are shown for each star in Fig.~\ref{fig:tess_aperture}. The spectral response function of the TESS is closer to the longer wavelength than the Johnson V-band \citep{tess}, therefore, the TESS light curves for these three systems were analyzed together with the only available NSVS \citep{nsvs} V-band light curves to obtain and compare the light contribution of the components in different passbands. The only available light curve data of TX Ari are taken from the NSVS catalog \citep{nsvs}.

Before starting the solution of light curves, the temperature of one of the components must be known. For this, either spectral or photometric color information is required. While atmospheric parameters can be accessed directly using spectral data, it is necessary to purify the colors from interstellar reddening (IS) in order to switch from photometric data to intrinsic colors and from there to temperature. Since all the systems we have considered are later than A0 spectral type, we cannot use the Q-method of \cite{Johnson} as with early-type stars. However, if we know the amount of interstellar reddening for the color indices we have considered, we can find the color excesses from here and switch to the true colors of the systems and from there to the temperature.

\begin{figure}[!t]
	\includegraphics[width=\columnwidth]{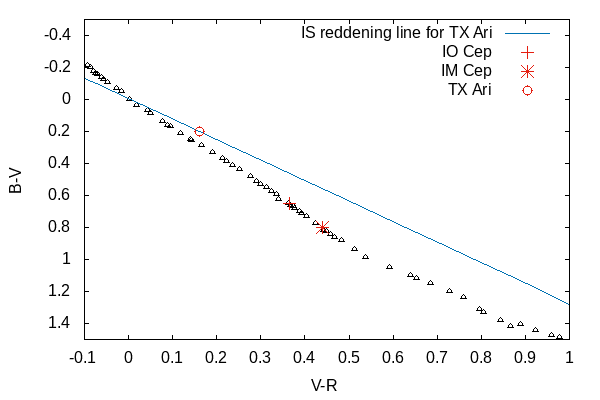}
	\caption{Location of the studied systems in color-color.}	
	\label{fig:color}
\end{figure}

For this reason, we obtained the $B-V$ and $V-R$ colors for each system with the help of conversion formulas described in GAIA DR2 documantation\footnote{https://gea.esac.esa.int/archive/documentation/GDR2/}, together with the $Bp$ and $Rp$ magnitudes of the systems in the GAIA \citep{Gaiateam2016} database. The color-color diagram created with the colors we have obtained is shown in Fig.\ref{fig:color}. The IS extinction slope in B--V vs V--R diagram is adopted from \cite{Savage} as $\frac{E(B-V)}{E(V-R)}=0.78$.

As can be seen from Fig.\ref{fig:color}, TX Ari moved away from the curve of unreddened main-sequence stars. However, since IO~Cep and IM~Cep are on this curve, either the amount of reddening is negligible or the amount of reddening is in the uncertainty box of the colors. For TX~Ari, moving along the direction of reddening towards the main-sequence, we obtain the unreddened colors for these systems. The observed colors of IO~Cep and IM~Cep are kept as they are. In Table~\ref{tab:colors_temperatures}, we give the photometric colors of the systems and the corresponding temperatures which are adopted from \cite{eker2020}.

\begin{table}[!t]\centering\small
	\caption{Intrinsic color indices, color excesses of program stars and corresponding temperatures.} \label{tab:colors_temperatures}
	\begin{tabular}{lccccc}
		\hline
		System & $B-V$ & $V-R$ & $E(B-V)$ & $E(V-R)$ & Temperature (K)\\
		\hline
		IO~Cep &0.65&0.37&0.0&0.0& 5770 \\
		IM~Cep &0.80&0.44&0.0&0.0& 5280 \\
		TX~Ari &0.20&0.16&0.20&0.16& 9200 \\
		\hline
	\end{tabular}
\end{table}

\begin{figure*}
	\begin{center}
		\begin{tabular}{cc}
			\resizebox{70mm}{!}{\includegraphics{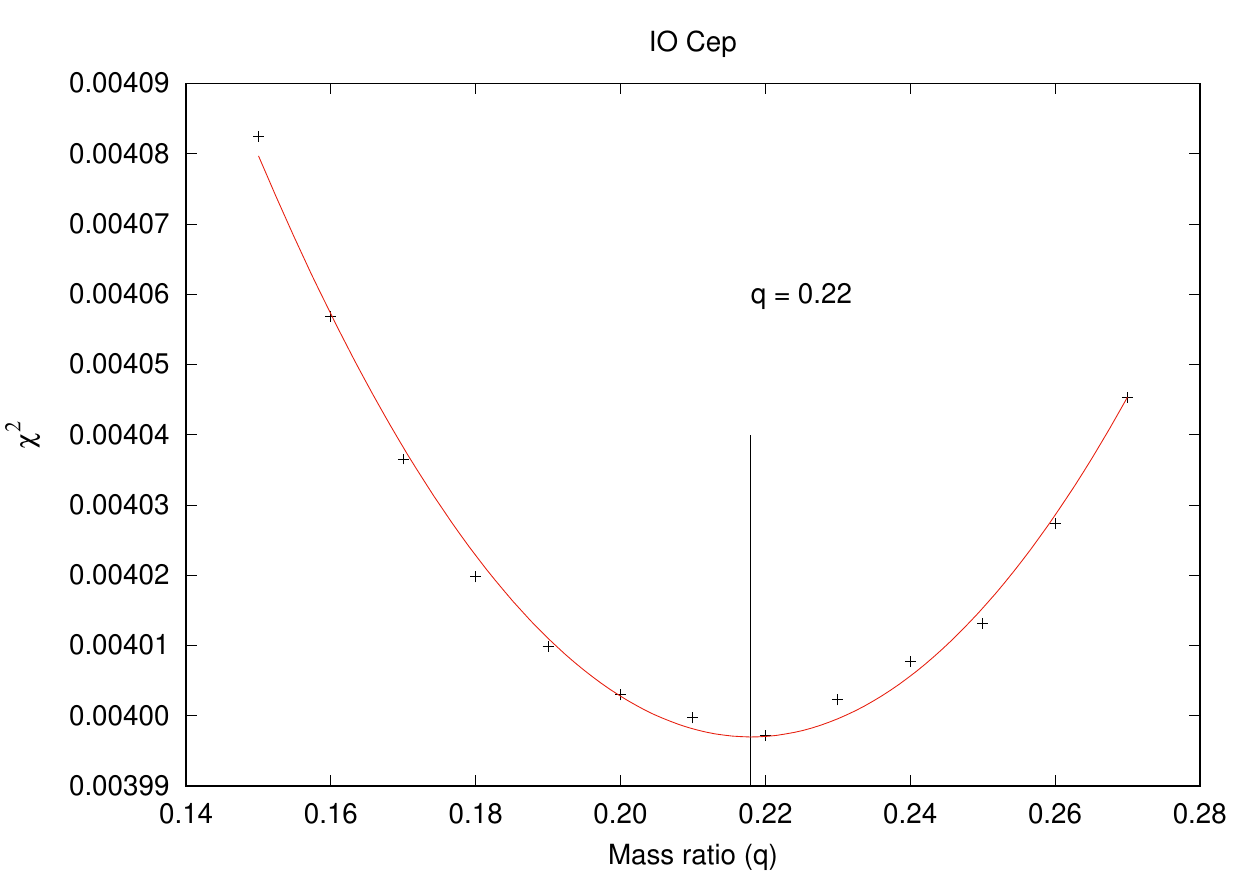}} & 
			\resizebox{70mm}{!}{\includegraphics{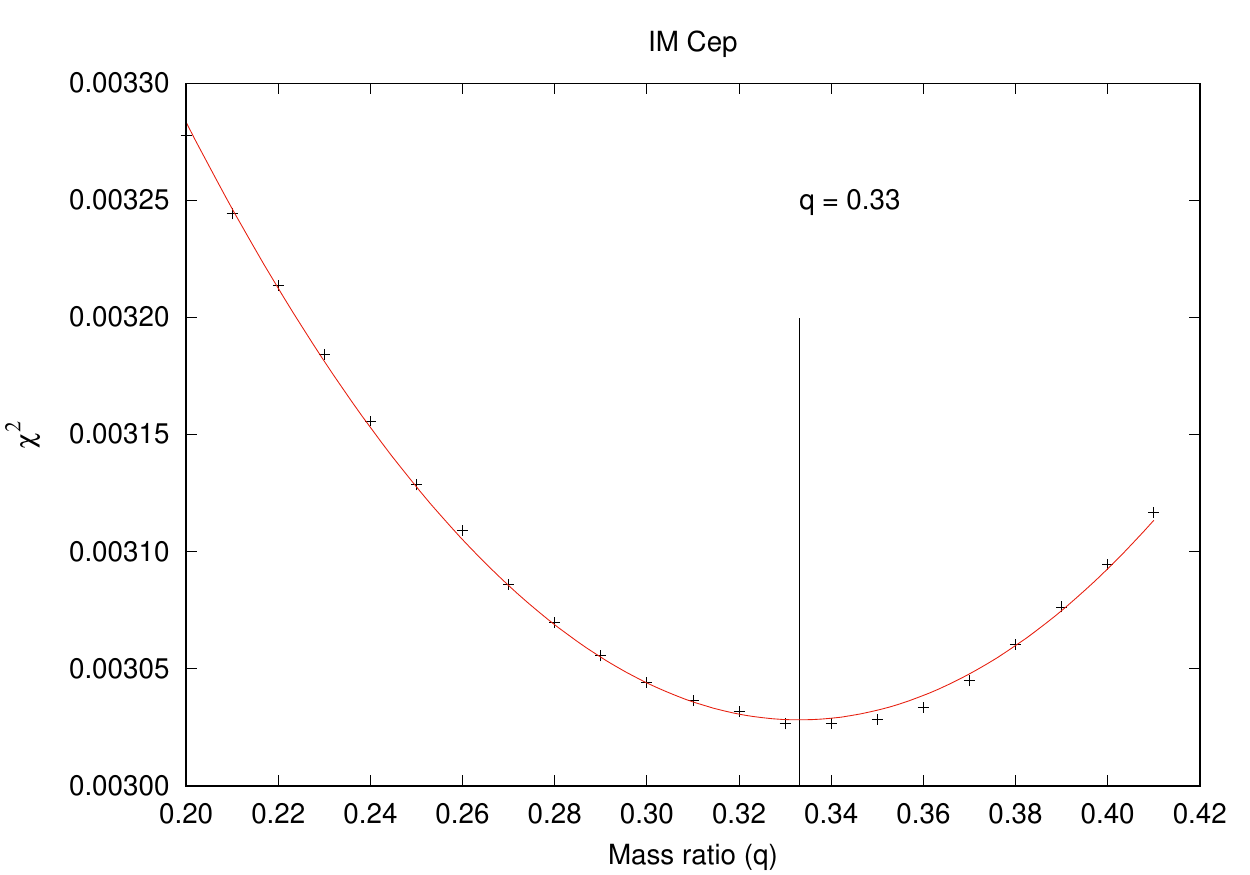}} \\
			\resizebox{70mm}{!}{\includegraphics{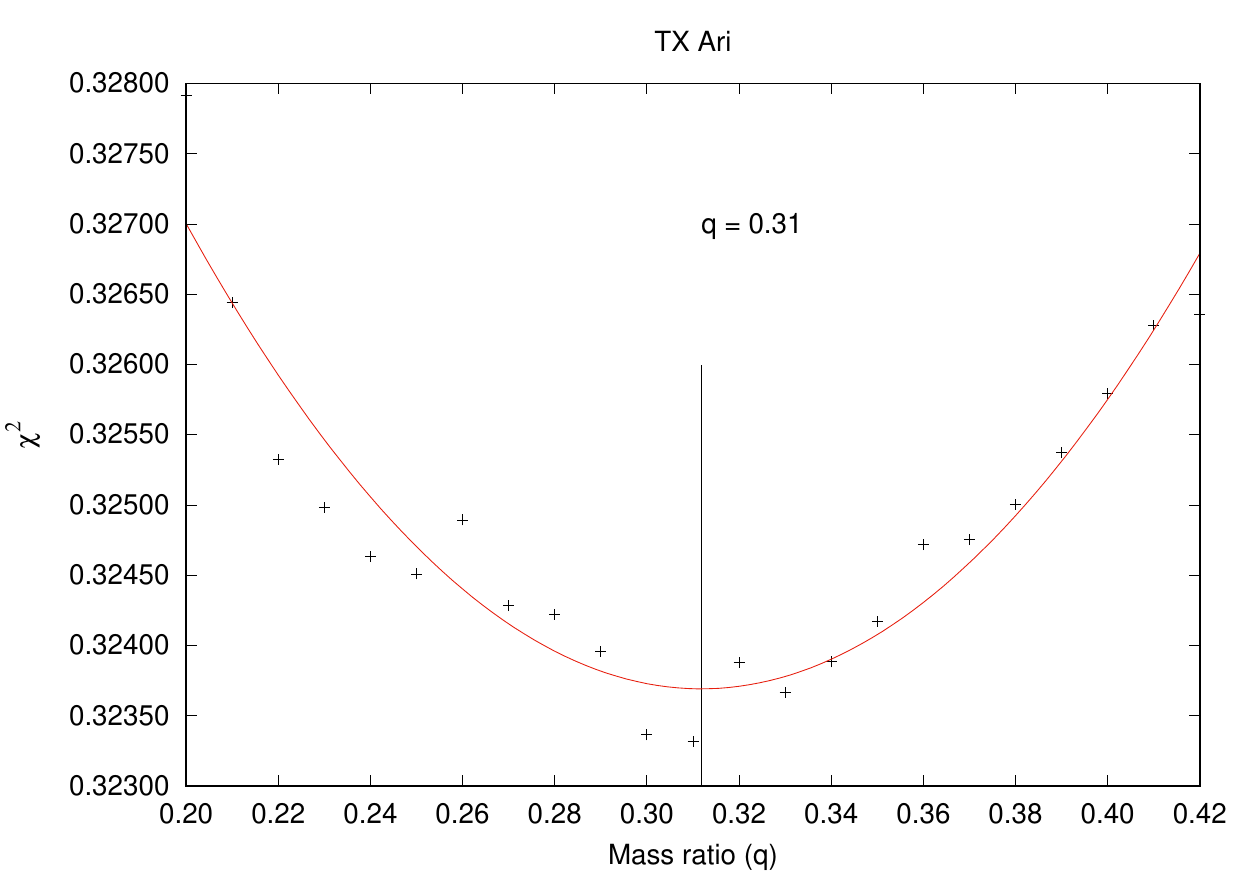}} & 
		\end{tabular}
		\caption{Variation of mass ratio with respect to $\chi^2$ in the model fits for each system.} \label{fig:qsearch}
	\end{center}
\end{figure*}

\begin{figure*}
	\begin{center}
		\begin{tabular}{cc}
			\resizebox{70mm}{!}{\includegraphics{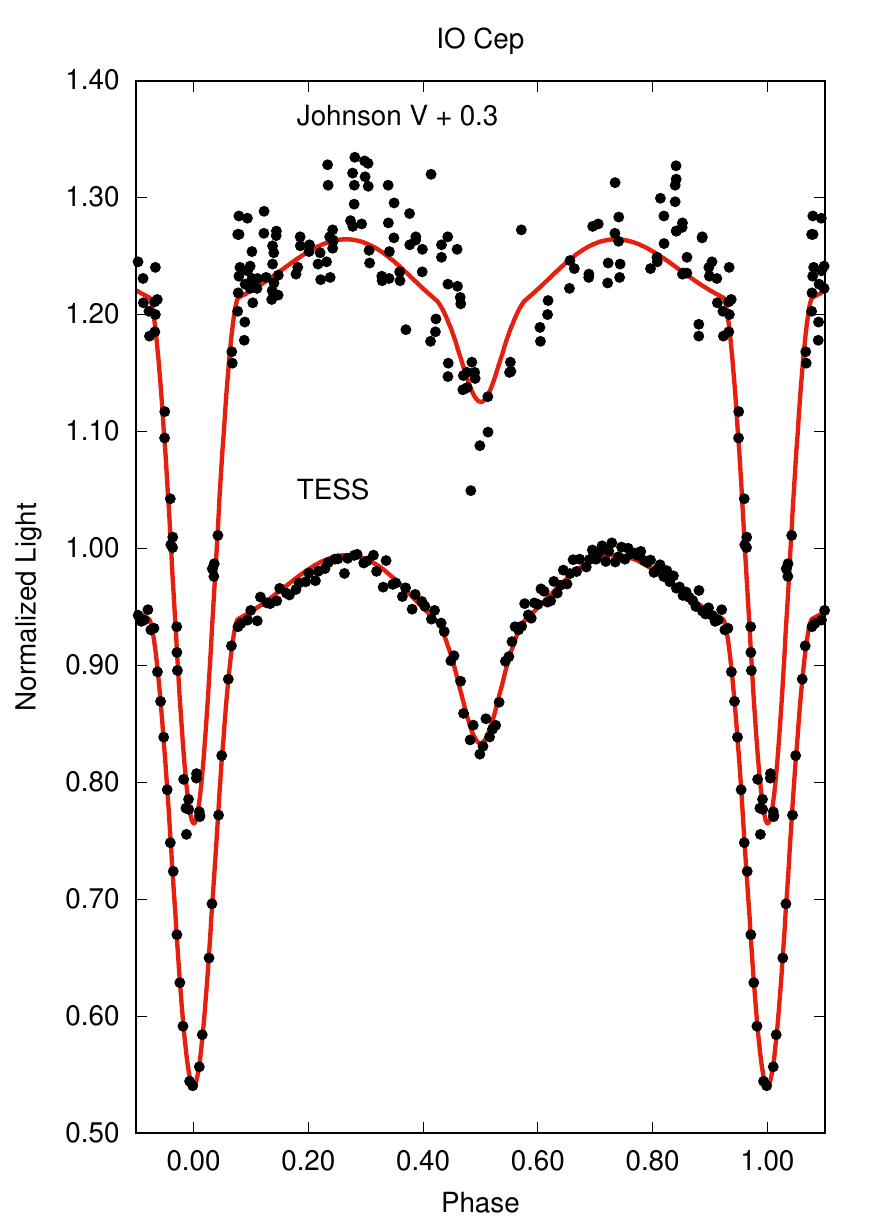}} & 
			\resizebox{70mm}{!}{\includegraphics{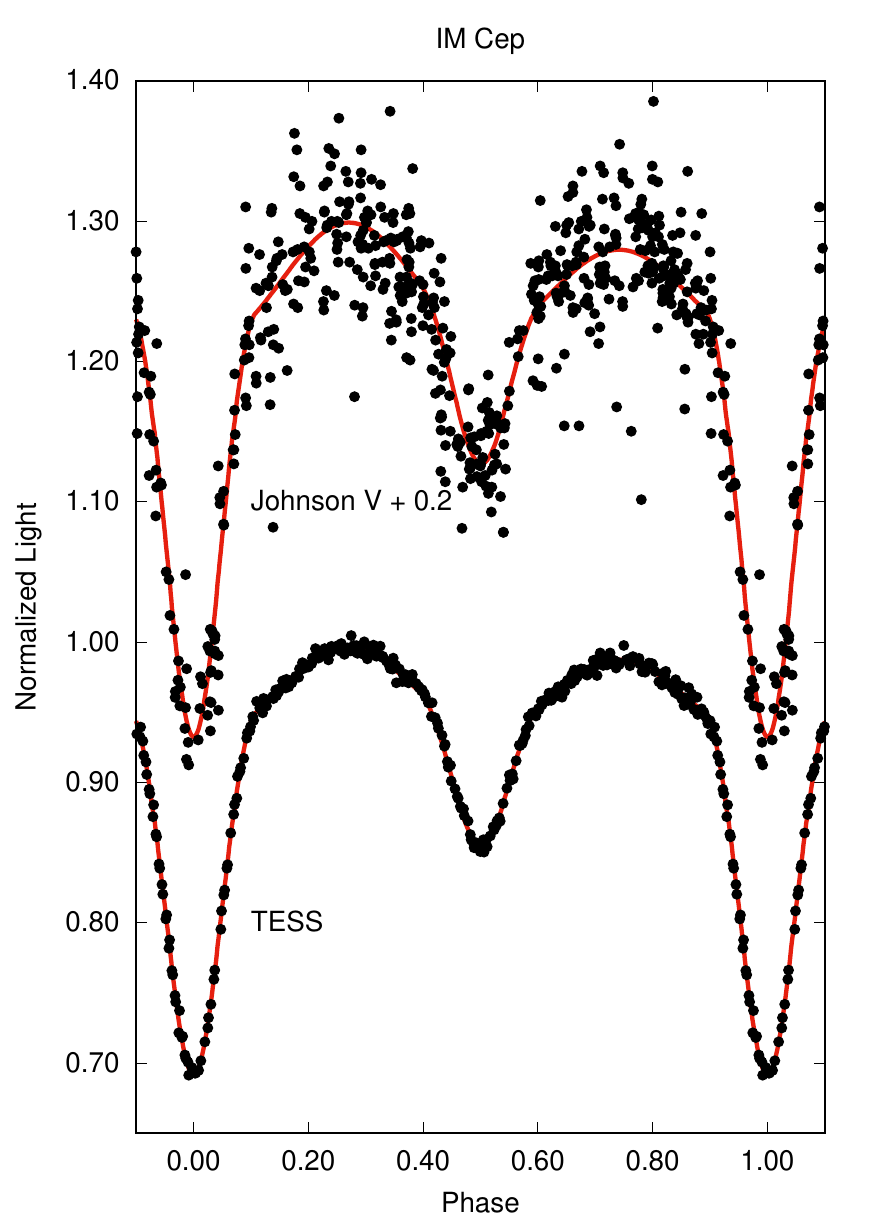}} \\
			\resizebox{70mm}{!}{\includegraphics{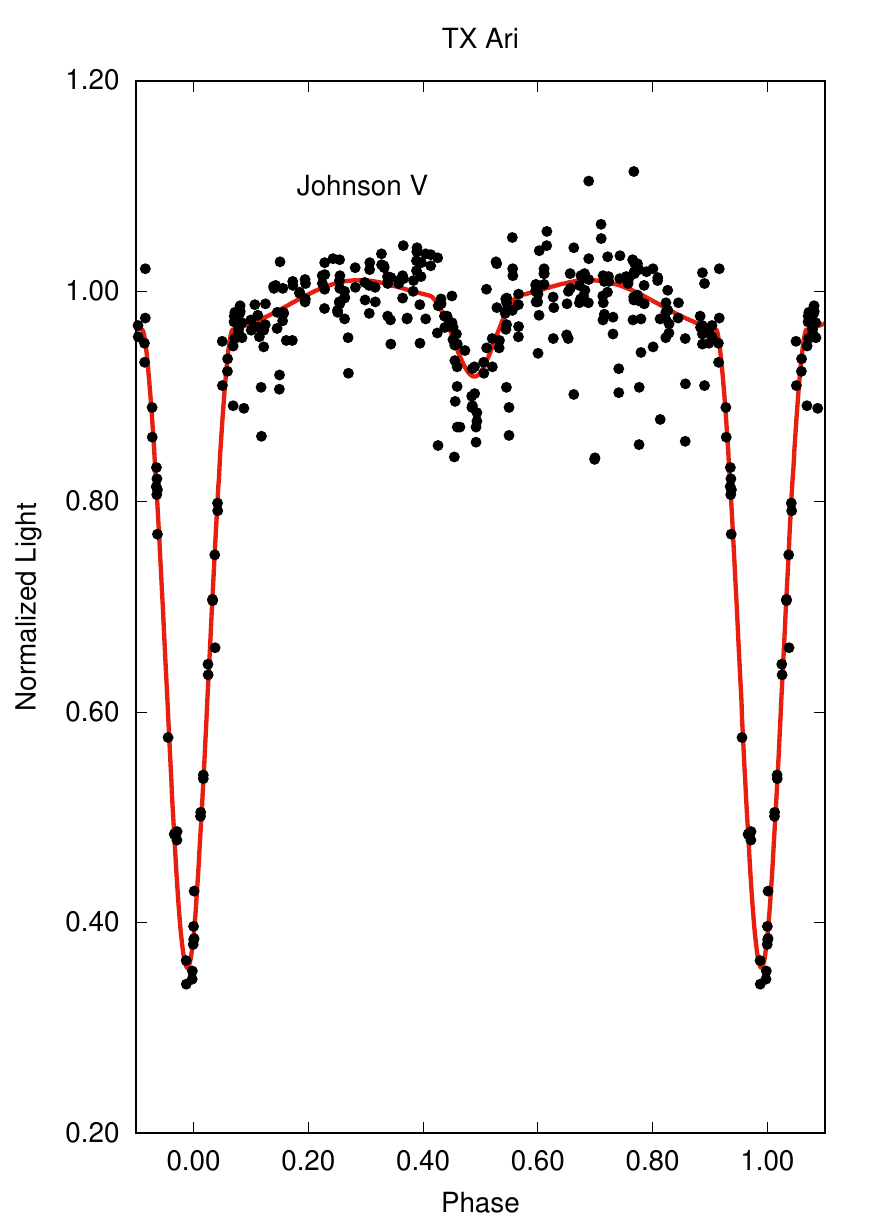}} & 
		\end{tabular}
		\caption{Light curve models for the systems.} \label{fig:lcmodels}
	\end{center}
\end{figure*}

\begin{figure*}
	\begin{center}
		\begin{tabular}{cc}
			\resizebox{70mm}{!}{\includegraphics{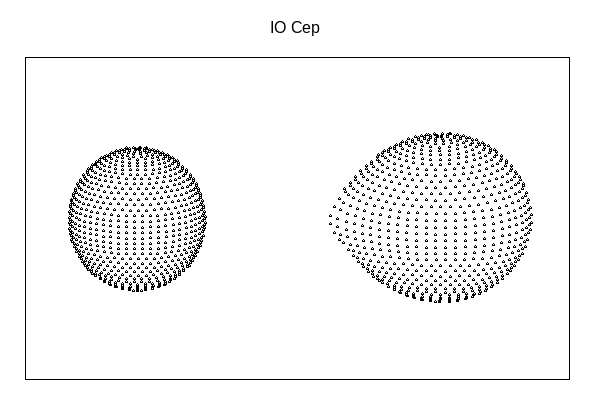}} & 
			\resizebox{70mm}{!}{\includegraphics{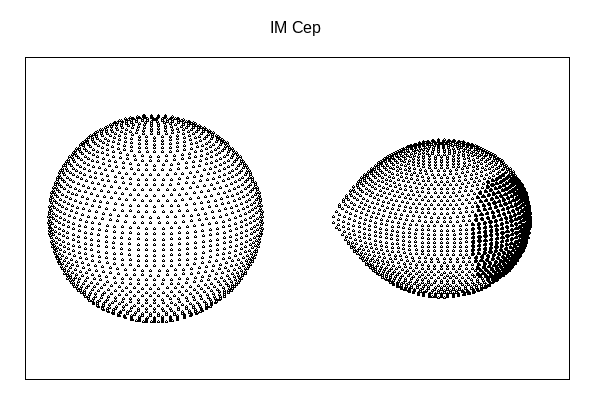}} \\
			\resizebox{70mm}{!}{\includegraphics{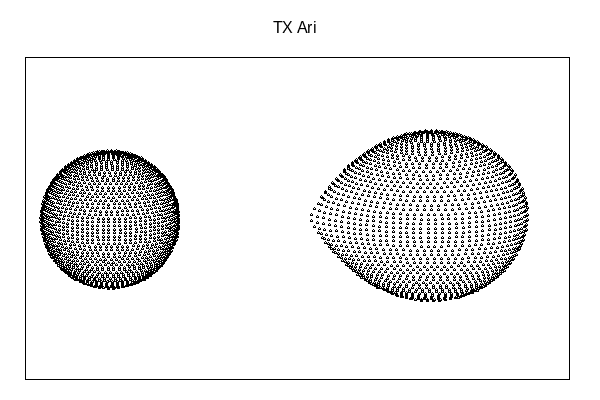}} & 
		\end{tabular}
		\caption{Roche surfaces of the component stars in each system.} \label{fig:roche}
	\end{center}
\end{figure*}

By obtaining the temperatures, mass ratio ($q$) distribution with respect to $\chi^2$ of model fits is established by analyzing the light curve of each system. The light curve solutions were performed using the well-known Wilson-Devinney analysis code \citep{WD_1971} under graphical interface of {\sc phoebe} \citep{phoebe_2011}. The code WD uses Roche equipotentials of the stars for the geometry of the system and the fitting is done with differential corrections algorithm. Before running differential corrections, the corresponding light curve with the input parameters such as inclination, equipotentials, secondary component temperature, light contribution of the primary component and spot parameters are calculated and plotted on observations. The radius of each star is a function of two parameters: its equipotential and the mass ratio of the system \citep{Kopal1978}. Therefore, adopting the best fitting mass ratio as the mass ratio of the system is a good method.  The $q-\chi^2$ graphs are given in Fig.\ref{fig:qsearch}. As can be seen from Fig.\ref{fig:qsearch}, a reliable mass-ratio for each system could be determined. Light curve models made with the obtained mass ratio and the corresponding Roche geometry of the systems are given in Figs. \ref{fig:lcmodels} and \ref{fig:roche}, respectively.

In addition to the level differences in the light curve quadratures ($\phi\sim$0.25 and 0.75) of IM Cep, asymmetrical structures draw attention. This was attributed to the secondary component being of the late spectral type, and the light curve was modeled with two spots, one at 150$^{\circ}$ and the other at 270$^{\circ}$. Both spots are at 90$^{\circ}$ latitude and their temperature is four percent cooler than the surface temperature of the secondary component.

Now, we have 3 systems with light curve parameters. However, all of these parameters are relative parameters and separation between components to be able to derive the absolute parameters such as mass and radius. One drawback might be that we don't have spectroscopic orbital elements. However, we can now make very accurate distance determinations with the parallax data from the GAIA mission. The following methodology could be used to compromise the absolute parameters of components using distance information. The total absolute magnitude $M_V$ of the system is obtained by using the distance and IS absorption ($A_v=3.1E(B-V)$) using the distance modulus. After obtaining the absolute magnitude of the components with the light contribution of each component (see Table~\ref{tab:absolutes}), the bolometric brightness of the components is obtained by using the bolometric corrections (BC) given by \cite{eker2020}. From the bolometric magnitudes, the absolute radius of each component is derived. Since we know the relative radii from the light curve analysis, the separation between components is found from the relationship $a = R_{absolute} / r_{relative}$. By finding the separation between the components and using the mass-ratio, the mass of each component and thus the total mass of each system can be derived. The absolute parameters of the components are derived and given in Table \ref{tab:absolutes}.

\begin{table}[!t]\centering\small
	\caption{Abasolute parameters of the systems. Errors quoted in brackets are internal errors of the parameters in the last digit. The parameters with no errors are not iterated.} \label{tab:absolutes}
	\begin{tabular}{lccc}
		\hline
		Parameter & IO~Cep & IM~Cep & TX~Ari\\
		\hline
		$P_{12}$ (days)& 1.2358109(6)& 0.9215840(2) & 2.691325(1) \\
		$m_{Vtotal}$ (mag) & 11.05 & 12.71 & 11.49 \\
		$A_V$ (mag) & 0 & 0 & 0.62 \\
		$d$ (pc) & 270(2) & 769(16) & 1246(53) \\
		$l_{1V}/l_{total}$ & 0.85(0.001) & 0.795(0.001) & 0.852(0.001) \\
		$l_{3TESS}/l_{total}$ & 0.09(0.01) & 0.133(0.001) & - \\
		$M_{v1}$      & 4.07(0.02)  & 3.53(0.05) & 0.57(0.71)\\
		$M_{v2}$      & 5.95(0.02)  & 5.00(0.06) & 2.47(0.72)\\
		$M_{bol1}$    & 4.05(0.02) & 3.37(0.05) & 0.72(0.71)\\
		$M_{bol2}$    & 5.29(0.02) & 4.47(0.06) & 3.14(0.72)\\
		$BC_{1,2}$ & -0.02, -0.66 & -0.16, -0.53 & -0.15, -0.68\\
		$i$ (deg) & 78.8(2) & 74.2(2) & 80.4(2) \\
		$T_{eff1}$ (K)    & 5770 & 5280 & 9200\\
		$T_{eff2}$ (K)    & 4380(40) & 4500(8) & 4400(50)\\
		$R_{1}$ ($R_\odot$) & 1.39(0.06) & 1.94(0.11) & 2.55(0.15)\\
		$R_{2}$ ($R_\odot$) & 1.36(0.08) & 1.05(0.08) & 3.61(0.34)\\
		$a$ ($R_\odot$)   & 5.73(0.29) & 6.32(0.54)  & 12.1(1.2) \\
		$M_{1}$ ($M_\odot$) & 1.36(0.21) & 2.74(0.57)& 2.21(0.54)\\
		$M_{2}$ ($M_\odot$) & 0.30(0.03) & 0.91(0.17) & 0.69(0.15)\\
		$q$ ($M_2/M_1$) & 0.22(0.01) & 0.33(0.01) & 0.31(0.01)\\
		\hline
	\end{tabular}
\end{table}

Now, the minimum mass of the distant components in each system can be derived by using Eq.\ref{eq:massfunction}. Moreover, masses for the case of coplanar orbit (same inclination of the wide and close binary orbits) can be derived. Mass estimates of the distant components are given Table~\ref{tab:third_mass} for each system.

\begin{table}[!t]\centering\small
	\caption{Estimated masses for the additional components in the studied systems. Errors in brackets refers to last digit.} \label{tab:third_mass}
	\begin{tabular}{lccc}
		\hline
		System & Component & $M_{min}~(M_\odot)$ & $M_{coplanar}~(M_\odot)$ \\
		\hline
		IO~Cep & 3 & 0.52(9) & 0.52(9) \\
		IM~Cep & 3 & 1.6(2)  & 1.7(2) \\
		TX~Ari & 3 & 2.9(3)  & 2.9(3) \\
		       & 4 & 2.7(3)  & 2.6(5) \\
		\hline
	\end{tabular}
\end{table}

\section{Discussion and Conclusion}
\label{sec:discussion}

The distance information provided by the Gaia satellite has enabled us to reach the physical properties of the components of three selected Algol-type binary stars. The current parameter accuracy is still not at the desired level of $<$ 1 percent, although, it offers a good opportunity to obtain a lower limit for the mass of physically bonded components around the systems under study.

Looking at the derived physical parameters of the components of the IO~Cep system, it is seen that the temperature of the primary component is close to the solar value while its mass and radius point to an earlier type system (F5-7). In parallel to this, the radius of the secondary component is larger than the radius of a star with 0.30 $M_\odot$ mass. These differences could not be explained by a single star evolution as there is mass transfer between the components in Algols, which separates the evolution of components from the evolution of single stars. It is clear that both components in IO~Cep evolved and increased their radius. The primary star is more massive for its temperature  (5770~K) while the secondary is less massive for its temperature (4380~K). The cyclic period change in the O--C diagram of the system is due to a distant companion as the theoretical model provides an eccentricity of $e=0.32$ which rules out orbital period modulation due to magnetic activity (see \cite{Applegate1992}). Using the total mass of the system (1.66 $M_\odot$) and the mass function of $f (m_3) = 0.029$ determined from the $O-C$ analysis, we conclude that there should be an additional component with a minimum mass of 0.52 $M_\odot$ revolving in a $\sim$36.9 yr orbit around the system. From the analysis of the TESS light curve, a 9 percent additional light contribution to the total light is noteworthy. This contribution is close to the light contribution of the second component of the close pair.

Concerning the features of the IM~Cep system, it is interesting that, despite its distance of 769 pc, there is no remarkable reddening in the direction of IM~Cep. It is worth noting that the mass of the primary component is high compared to its temperature and the radius of the secondary component is larger than the radius of a star with similar mass. This is what we are used to see in an Algol-type system transferring mass between components. While the primary component increases its mass during the mass transfer, it does not increase its luminosity since the gained mass does not change the energy production in the core of the star. Therefore the primary looks underluminous in the H-R diagram. Nevertheless, the mass loosing secondary star does not change its luminosity as reducing its mass which causes it to be seen as an overluminous star in the H-R diagram. This is exactly the case of IM~Cep, the primary star has increased its mass to 2.74 M$_\odot$, a mass of an B9-type main-sequence (MS) star, and evolved to a radius of 1.94 R$_\odot$, an F0-type MS star, while cooling down to a temperature of a K4 spectral type star. The mass loosing secondary star follows similar dramatic evolution scenario. It has now a mass of an G8-type star with a solar-like radius while cooling down to a K5 spectral type star. In addition to its evolutionary scenario, it is noteworthy that the light curve analysis yielded the secondary component as an active star. The asymmetrical structure in the light curve and the level differences in the maximum phases can only be modeled with two large spots on the surface of the secondary component. If the cyclic variation in the $O-C$ diagram (see Fig.~\ref{fig:imcep}) is not due to magnetic cycle, using the total mass of the system (3.65 $M_\odot$) and the mass function of $f (m_3) = 0.144$ determined from the $O-C$ analysis, we conclude that there should be an additional component with a minimum mass of 1.6 $M_\odot$ revolving in a $\sim84$ yr orbit around the system. The 13 percent light contribution of the third component in the TESS light curve strengthens the possibility that the source of the cyclic change in the $O-C$ curve is due to the third body. It should be noted here that the third component contributes more than the secondary component of the close pair which is in agreement with the difference in their masses. Nevertheless, unfortunatelly, the possibility of the source of the cyclic change in the O--C curve could be due to magnetic activity could not be ruled out as there is no historical brightness data of the system to check for the Applegate mechanism \citep{Applegate1992}.

Perhaps the most interesting of the systems studied is TX~Ari. The extinction amount in the V-band of the 1246 pc away system is 0$^m$.62. Since it has not been observed by the TESS satellite yet, it was not possible to compare the light contribution rates of the components in two different bands. In TX~Ari, the radius of the primary component appears to be compatible with its mass, while the radius of the secondary component is very large compared to its mass. It is interesting to note that the $Q$-parameter in the $O-C$ diagram which is responsible for the mass transfer from the less massive secondary component to the primary component is zero. A small amount of mass transfer may be present in the system however it could be suppressed by the effects of two sinusoidals in the $O-C$ diagram. Two cyclic changes stand out in the $O-C$ analysis of the system. First cyclic change is modelled with an eccentric LTT orbit with an orbital period of 70 yrs and projected semi-major axis of 14.4 AU indicating a mass function of $f(m_3) = 0.61$. This mass function propose a minimum mass of 2.7 M$_\odot$ third body in the system. Second cyclic change is indicates a zero eccentric orbit with a period of 33 yrs which can be due to magnetic cycle of the late type 0.7 M$_\odot$ secondary component. If it is not the magnetic activity, the system contains another distant component with a minimum mass of 2.6 M$_\odot$, which makes the the system hierarchical quadropule. Another feature of such hierarchical multiple systems is the mean-motion resonance of the distant companions \citep[e.g.][]{Yuan2016}. Orbital periods of distant components of TX~Ari are $P_3=70\pm6.0$ yrs and $P_4=32.7\pm2.6$ yrs, which may indicate a possible 2:1 mean-motion resonance within the uncertainty of the periods. TX~Ari deserves further multiband photometric and high resolution spectroscopic observations to reveal the nature of the system.

\begin{figure*}
	\begin{center}
		\begin{tabular}{c}
			\resizebox{120mm}{!}{\includegraphics{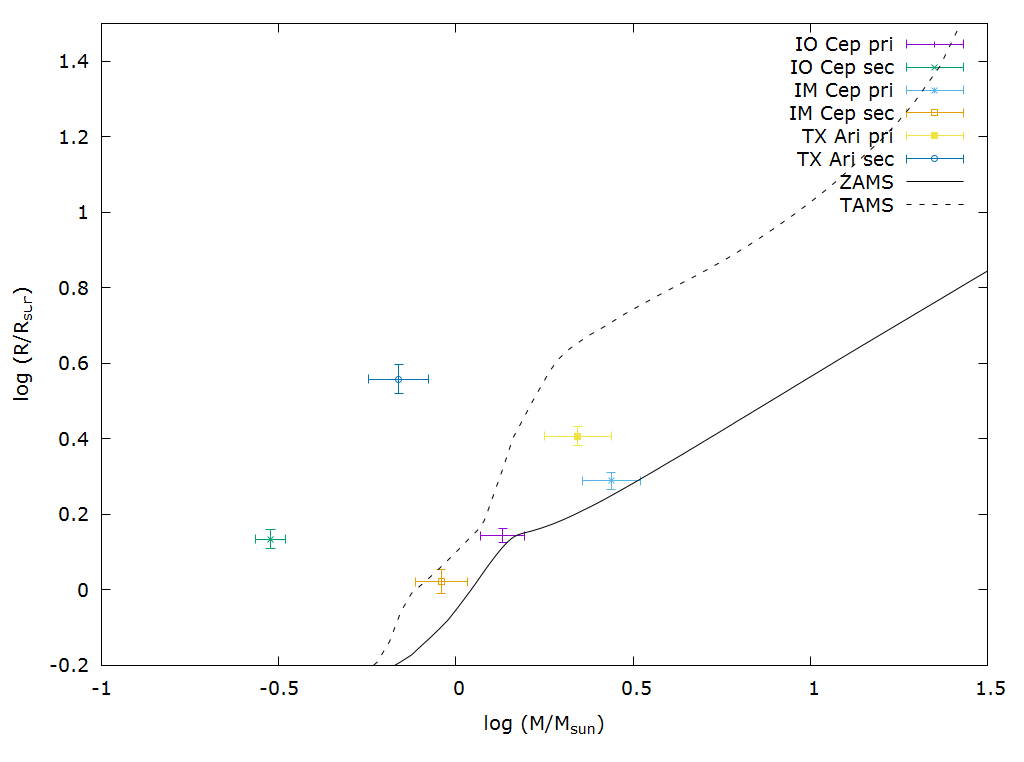}}
		\end{tabular}
		\caption{Location of component stars of IO Cep, IM Cep and TX Ari in the $log M - log R$ plane. ZAMS and TAMS lines are shown with solid and dashed lines, respectively.} \label{fig:logMlogR}
	\end{center}
\end{figure*}

The evolution of semidetached binaries on the basis of their angular momentum change are very well described by \cite{Ibanoglu2006}. Three cases of mass transfer occurs in semidetached Algols: Case A, B and C. In case A, the transfer occurs during core hydrogen burning phase, in Case B, during shell hydrogen burning phase and in Case C, the mass transfer occurs when the star has exhausted helium in its core. In Fig.\ref{fig:logMlogR}, we show the positions of the component stars of three systems with ZAMS (Zero Age Main Sequence) and TAMS (Terminal Age Main Sequence) curves in $log M-log R$ plane. As can be seen from Fig.\ref{fig:logMlogR}, all studied Algol systems have experienced mass reversal. While the primary components of the systems are in the main-sequence, secondary components of IO~Cep and TX~Ari are clearly deparated from the main-sequence. Nevertheless, secondary component of IM~Cep seems to be close to the TAMS line within the uncertainty box. If secondary component of IM~Cep is still in the main-sequence it have experienced mass reversal during the main-sequence life and the donor secondary component continues transferring mass in the main-sequence. IO~Cep and TX~Ari clearly perform Case-B type mass transfer.

\section*{References}

\bibliography{bibfile}

\end{document}